**Quantum Algorithm Processor For Finding Exact Divisors**
Professor J. R. Burger

*Summary* – Wiring diagrams are given for a quantum algorithm processor in CMOS to compute, in parallel, all divisors of an n-bit integer. Lines required in a wiring diagram are proportional to n. Execution time is proportional to $n^2$.

**Introduction**
The goal is to find factors of a given number N with n bits as required for various purposes. This may be done in a quantum algorithm processor fabricated in CMOS as originally promoted by the author [1]. Each register may execute in parallel the instructions in a wiring diagram to accomplish division by integers from 2 to √N. This amounts to dividing by all n/2 bit divisors (for n even). A zero remainder is indicative of exact factors. Nonrestoring division, the kind that does not use zeros in the quotient, was considered and found not to work as well for detecting zero remainders. This section suggests an appropriate wiring diagram that accomplishes restoring division. Later, to illustrate a tradeoff, division by successive subtraction is considered.

The goal differs from prime factorization. Prime factorization would initialize all divisors to be prime numbers instead of integers*

Central to the problem being solved is the reversible adder. Figure A1 shows a symbol for a reversible adder.

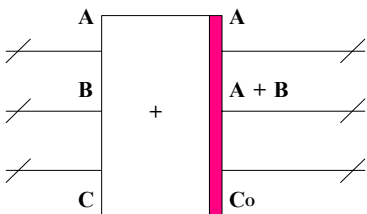
Figure A1. Reversible adder symbol

Logic for the adder is available [2]. There is no carry-in to this version of the adder, but there is a carry-out. So each bus has the same number of lines.

**Restoring Division**
As a reminder of how this works, consider 1111 divided by 11 as in Figure A2. A sign bit is appended to ensure 3 place arithmetic as appropriate for a 2-bit divisor.

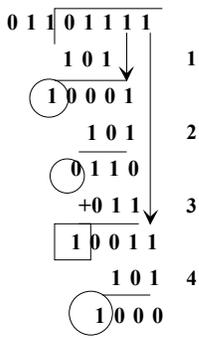

Figure A2   Restoring division example

The 2s complement of the divisor 011 is 101.  This is added to the leftmost bits of the dividend at step 1.  If the carry is 1, the result of the operation is positive.  Bring down the next bit of the dividend and add 101 again as in step 2.  In this step the carry is 0, so the result is negative and must be restored.  Add 011 as in step 3. The carry out of the addition is always 1 because the result is always positive.  Then bring down the last bit of the dividend.  Add 101 as in step 4.  The remainder is 000 in this case.  The quotient is built from the circled carries to be 101.  That is, 15 divided by 3 equals 5, or  Mod(15, 3) = 0.

For larger numbers there have to be leading zeros to perform fixed point division with a fixed number of bits.  Such details are omitted below because they are easily included, and because they distract from the process.  A 4-bit dividend is imagined, but the design is extendable to n bits.

**Twos Complement Generator**
Division in the diagrams below uses 2-bit divisors, but a sign bit is required, so actually there are three bits.  Figure A3 shows inputs and outputs.

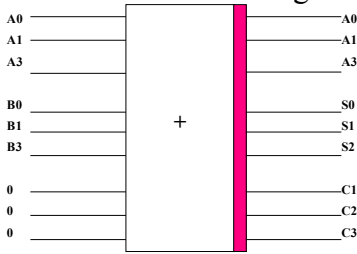

Figure A3.  Three-bit adder

The twos complement of the divisor is obtained by individually complementing each bit of the divisor and then adding one.  The plan is shown in Figure A4.

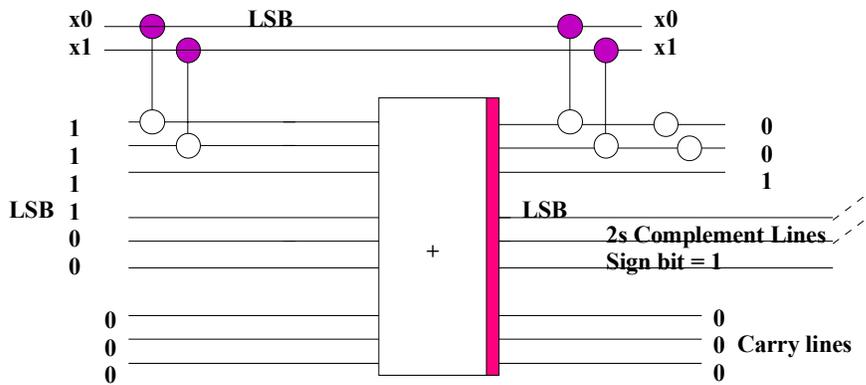

Figure A4. Twos complement generator

Note that the A outputs of the adder are zeroed. The divisor and its complement will be carried through each stage of the division.

**Conditional Subtract-Add Stage (CSA)**
To convey the plan clearly, a four bit dividend with sign bit $0y_3y_2y_1y_0$, for example, 01111, is thought of as being divided by a two bit divisor with sign bit $0x_1x_0$, for example, 011. Generalization to larger numbers is messy but uncomplicated. In restoring division there is a certain partial dividend. The divisor is subtracted conditionally from the partial dividend. Subtraction is accomplished by adding the 2s complement. If the result of the subtraction is negative then the divisor is added back to the result to restore the partial dividend, and a 0 is appended to the quotient. If the result is positive, then there is no restoration, and a 1 is appended to the quotient. No restoration is equivalent to adding zeros. Adding zeros works via the wiring diagram logic below. Figure A5 is the plan for conditional subtract-add.

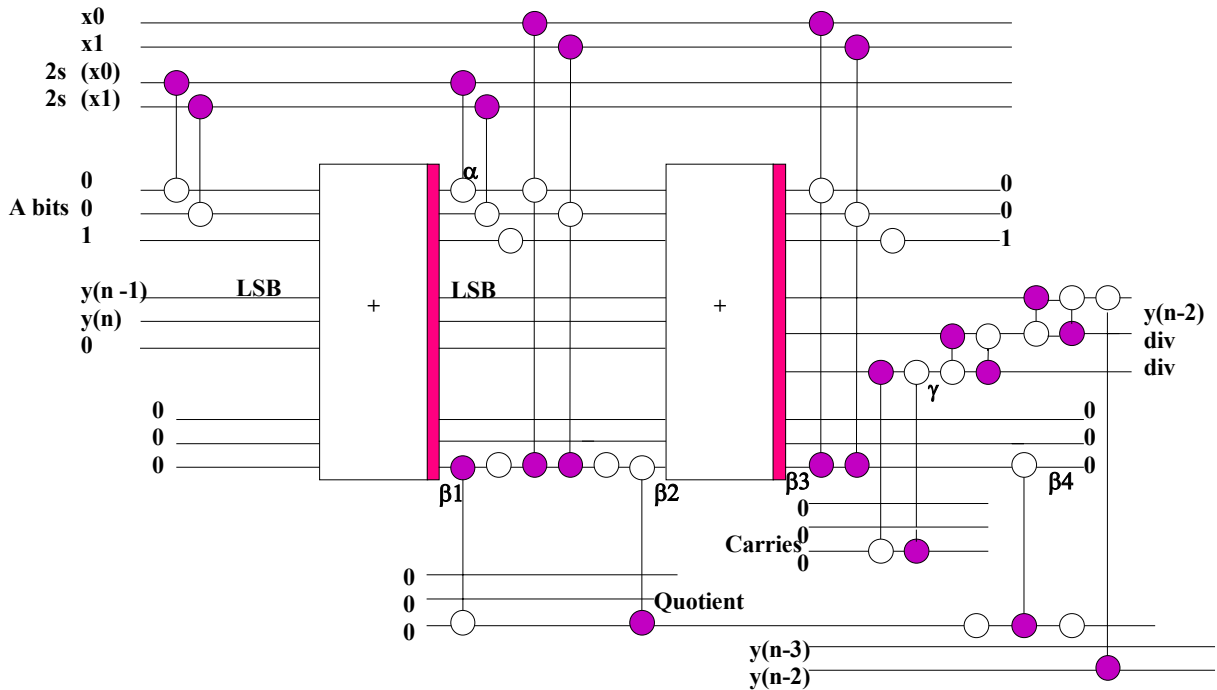

Figure A5   Conditional subtract-add (CSA) stage

The A bits in Figure A4 are used to hold the 2s-complement of the divisor. Note that it is unnecessary to have the addend bits adjacent. They communicate with vertical lines, that is, inter register buses in a CMOS implementation.

The carry out process at $\beta 1$ is a little tricky. If the carry out at $\beta 1$ is 1 then a 1 is placed in the quotient lines in the most significant bit (MSB) place. At $\alpha$ the A-lines are zeroed. Zeros are added to the partial dividend. Appropriate inverters restore the 1 at $\beta 1$ to 0 at $\beta 2$. It stays 0 at $\beta 3$ since only 0s are added. It remains 0 at $\beta 4$.

If the carry out is zero, the divisor is added to restore the partial dividend. If the carry out is 0 at $\beta 1$, it will be 0 at $\beta 2$. But the carry out at $\beta 3$ will be 1. This is inverted to 0 at $\beta 4$.

For division, there is a shift down at $\gamma$ to prepare for the next stage. The carry out is saved in lines labeled Carries. The least significant bit (LSB) of the result is moved to the next most significant place. The next bit of the dividend y(n-2) is brought into the LSB place. This prepares the stage to drive similar stages. Subsequently the process is repeated until all dividend bits are processed. The main differences are that the quotient, carry and dividend lines change for each iteration. Figure A6 shows how a zero remainder is detected.

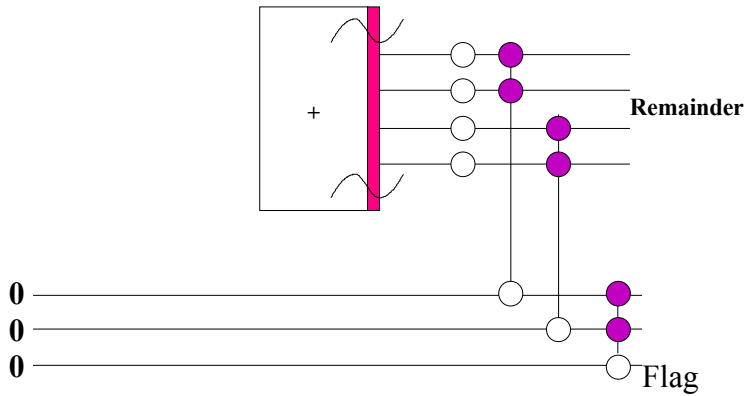

Figure A6   Last step

For n even, and n/2 remainder bits, the number of lines to detect zero remainder is $n/4 + n/2 + \ldots 1 = n/2 - 1$. If the remainder is zero, the Flag will be one.

**Performance Analysis**

There are two major considerations when using CMOS: size of memory and speed of memory.

**Size** – The size of the dividend without the sign bit is n  For the purpose of simplifying; let n be an even number. Then divisors (without the sign bit) will be n/2 bits. The divisor uses n/2 lines that are carried through the entire diagram. The 2s complement uses another n/2 lines that are carried through the entire diagram.

The 2s-complement generator uses 3 [n/2 + 1] lines. These are used again in CSA stages. The dividend, except for the two most significant bits going directly to the first stage, uses n-2 lines that are carried through the entire operation. The number of quotient lines is n-1. The number of carry lines also is n-1. Finally, the test for zero remainder needs (n/2)-1 extra lines. Approximately, the number of lines (BITS per CMOS register) required is:
BITS = n/2 + n/2 + 3[n/2 + 1] + n-2 + n-1 + n-1 + (n/2)-1 ≈ 6 n

This is not optimized yet. But it compares to the 4n result [2] or the 5n result [3].

**Time to complete** – Operations can be accounted as follows: The adder circuit for n/2+1 bits uses about 4n+14 steps or gates. Twos complementing uses this much plus 3n/2 steps. There will be n-1 CSA stages. CSA uses two adders plus about 3n+7 steps. Finalizing uses n/2-1 steps. The net number of operations (OPS) is approximately
OPS = 4n+14 + 3n/2 + (n-1)[2(4n+14) + 3n+7] + n/2-1 ≈ 11 $n^2$

This is much better than the order of $n^3$ result [2,3].

**Taking advantage of irreversible logic**

A feature of CMOS technology is that, aside from being deterministic, it is capable of being irreversible. For the purpose of this paper, being irreversible means that a bit can be reset to zero. Symbols for this are suggested in Figure A7.

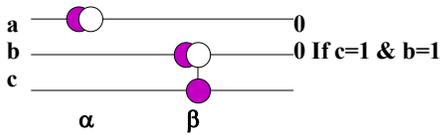

Figure A7 Irreversible operations available

At α the a-line is zeroed if a = 1. If a = 0, it stays zero. At β the b-line is zeroed only if c=1 while b=1. If b=0, it stays zero. If c=0, b is unchanged. These are irreversible, since for example, once zeroed, the original value of a cannot be reconstructed. Operations as above are readily available in the author's quantum algorithm processor using CMOS. Algorithms that use zeroing go well beyond quantum computers, since quantum systems must be reversible.

Consider the CSA stage as modified in Figure A8 to use irreversible operations.

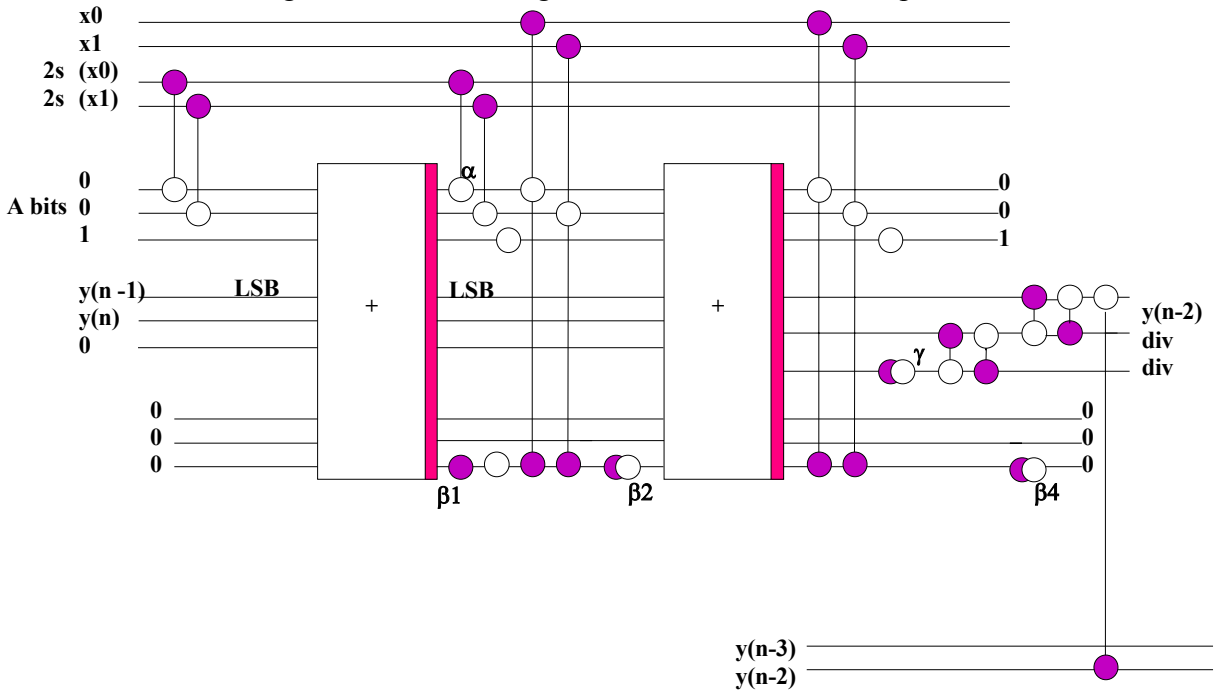

Figure A8   Simplification using irreversible operations

The quotient bits at β1 no longer need to be saved because the carry can be irreversibly reset to zero at β2 and β4. Similarly, the carries at γ no longer need to be saved because the MSB of the partial dividend can be reset to zero. The number of lines is reduced to:
BITS = n/2 + n/2 + 3[n/2 + 1] + n-2 + (n/2)-1 ≈ 4 n

This is the best so far. A further reduction of n/2 might result by designing an improved adder that also subtracts, so that a 2s-complement does not need to be carried through (Not done here).

**Division By Successive Subtraction (SS)**
This illustrates a tradeoff, using fewer lines at the expense of more operations. Sign bits can be ignored. An n-bit 2s complement of the divisor is added to an n-bit dividend until the carry out goes to 0. Finally, if the divisor is an exact factor of the dividend, the remainder at this point must equal the 2s complement of the divisor. Figure A9 suggests an implementation.

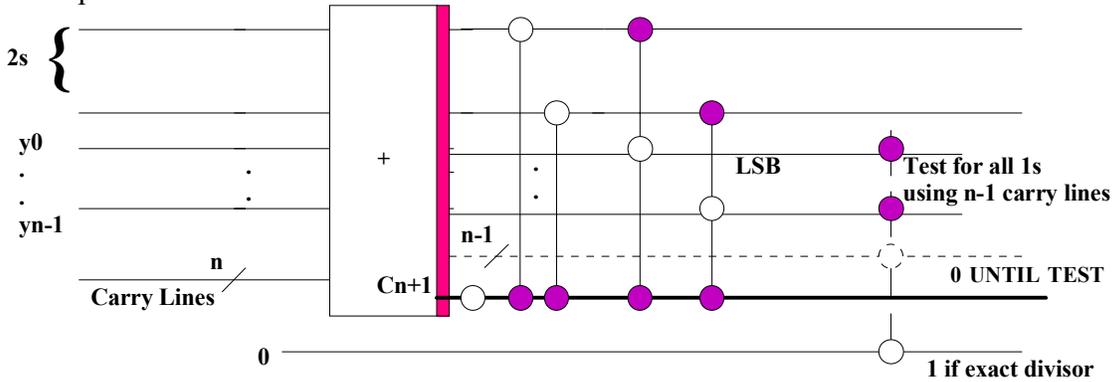

Figure A9   Successive subtraction stage with testing

The 2s complement is generated as before using leading ones to make n bits. Twos complement subtraction repeats until the carry out changes to 0, at which time a test is initiated. First, the carry out is converted to one. The objective at this point is to determine if the remainder equals the 2s complement, indicating an exact divisor. Second, 2s complement lines are inverted so that 0 becomes 1. Third, each '1' conditionally flips bits in the remainder. The remainder will be all ones <u>only</u> if there is a match. Subsequently a standard test is performed to determine if the converted remainder is all ones (as in Figure A6). This test uses the scratchpad carry bits, all n-1 of them. Finally, a flag bit is set if there is a match. Note that the left-overs will proceed through the system, but that the results have no meaning in this paper.

**SS Size** – Successive subtraction uses 3n lines plus a flag line. This is better than restoring division as above.
BITS ≈ 3n

**SS Time** – There will have to be about $2^{n-1}$ subtractions in this system (in a worst case) even though most numbers do not require this many. Each subtraction takes about 8n+6 operations. The testing uses about another 3n operations. Thus
OPS ≈ 11n $2^{n-1}$

Because of the exponential growth in delay with size, successive subtraction is a poor choice for larger applications.

In contrast, nonrestoring division for a 32 bit dividend can use 128 bit registers. A 16 bit subregister holds the divisors. All possible integer divisors fit into 64K (or 65,536) registers. This is possible physically for n not too large. Optimization and the expected exponential improvement in technology each year will push n into the hundreds and beyond.

Assuming a modest one microsecond (1 μs) per operation, delay for 11 $n^2$ operations is about 10 ms per 32 bit dividend,. In practice, CMOS clocks are higher than one megahertz, so delay is less of an issue.

**Conclusion**
Nonrestoring, restoring, and successive subtraction division were considered for finding simultaneously all factors of a given integer. Nonrestoring division as commonly practiced cannot easily detect a zero remainder. Restoring division was found to work in CMOS using 4n lines in a wiring diagram, and approximately $11n^2$ operations. Successive subtraction works with 3n lines and roughly $11 n 2^{n-1}$ operations. Restoring division appears to be best for larger applications.

With appropriate engineering help, a quantum algorithm processor in CMOS is readily available. CMOS is deterministic and supportive of irreversible operations, a real advantage in practice. A quantum algorithm processor in CMOS typically uses $2^m$ registers to perform calculations in parallel on all combinations of m bits, as mentioned above. It is potentially more efficient that any classical computer that uses fewer than $2^m$ parallel processors.

**References**

[1]    John Robert Burger, Data-stationary architecture to execute quantum algorithms classically, http://arXiv.org/abs/cs.arch /cs.AR/0412040, 9 Dec 2004.

[2]    V. Vedral, A. Barenco, and A. Ekert, Quantum networks for elementary arithmetic operations, Physical Review A, Vol. 54, No. 1, pp. 147-153, July 1996.

[3]    D. Beckman, A. Chari, S. Devabhaktuni, and J. Preskill, Efficient networks for quantum factoring, Physical Review A, Vol. 54, No. 2, pp. 1034-1063, August 1996.

**\*Integer factorization From Wikipedia, the free encyclopedia.**
In number theory, the integer factorization problem is the problem of finding a non-trivial divisor of a composite number; for example, given a number like 91, the challenge is to find a number such as 7 which divides it. When the numbers are very large, no efficient algorithm is known; a recent effort which factored a 200 digit number (RSA-200) took eighteen months and used over half a century of computer time. The supposed difficulty of this problem is at the heart of certain algorithms in cryptography such as RSA. Many areas of mathematics and computer science have been brought to bear on the problem, including elliptic curves, algebraic number theory, and quantum computing.